\documentclass[twocolumn,showpacs]{revtex4-1}
\usepackage{graphicx}
\usepackage{amsmath}
\usepackage{amssymb,amsthm}
\graphicspath{{pict/}{}}
\usepackage{bm}

\newcounter{Fig}

\newcommand{\be}{\begin{equation}}
\newcommand{\ee}{\end{equation}}

\begin{document}

\title{Surface bound states in the continuum}

\author{Mario I. Molina$^{1}$, Andrey E. Miroshnichenko$^{2}$ and Yuri S. Kivshar$^{2}$}

\affiliation{$^{1}$Departamento de F\'{\i}sica, Facultad de Ciencias, Universidad de Chile, Santiago\\
and Center for Optics and Photonics (CEFOP), Casilla 4016, Concepci\'{o}n, Chile\\
$^{2}$ Nonlinear Physics Centre, Research School of Physics and Engineering,
Australian National University, Canberra ACT 0200, Australia}

\begin{abstract}
We introduce a novel concept of surface bound states in the continuum, i.e. surface modes embedded into the linear spectral band of a discrete lattice. We suggest an efficient method for creating such surface modes and the local bounded potential necessary to support such embedded 
modes. We demonstrate that the embedded modes are structural stable, and the position of their eigenvalues inside the spectral band
can be tuned continuously by adding weak nonlinearity.
\end{abstract}

\maketitle

Soon after the emergence of quantum mechanics, von Neumann and Wigner suggested~\cite{wigner} that certain potentials could support spatially localized states within the continuum spectrum, i.e. bound states with the energies above the potential barriers. Since 1929, when this remarkable proposal was published, the bound states in the continuum were regarded as a mathematical curiosity, even though such potentials were suggested to occur in certain atomic and molecular systems~\cite{stillenger,shastry}. The subsequent experiments with semiconductor heterostructures provided the direct observation of electronic bound states above a potential well localized by Bragg reflections~\cite{capasso}.

In addition to the physics of electronic structures and quantum dot systems, this topic attracted a lot of attention
in optics~\cite{shabanov,bulgakov,moiseev}, where it was very recently shown that the optical bound states can be generated
in an optical waveguide array by decoupling from the continuum by virtue of symmetry only~\cite{segev}.

In this Letter, we extent the pioneering concept of von Neumann and Wigner~\cite{wigner} into two novel directions. First, we demonstrate that
the bound states can exist in systems of a semi-infinite extent as {\em surface bound states in the continuum}. We suggest and implement
a novel method for creating square-integrable, discrete surface modes embedded into a linear spectrum.  Such surface modes can be regarded
as a novel type of localized surface Tamm-like states with energies in the continuum (i.e. ``embedded Tamm modes''). Second, we study 
the properties of such embedded states in the presence of nonlinearity and demonstrate that the mode location inside the band 
can be tuned continuously by changing the mode amplitude.  Importantly, our modes appear in entirely asymmetric systems 
and therefore, cannot be reduced to the bound states in infinite systems where the decoupling
from the continuum occurs due to the symmetry conditions.

\begin{figure}[t]
\centering
\includegraphics[width=0.45\textwidth]{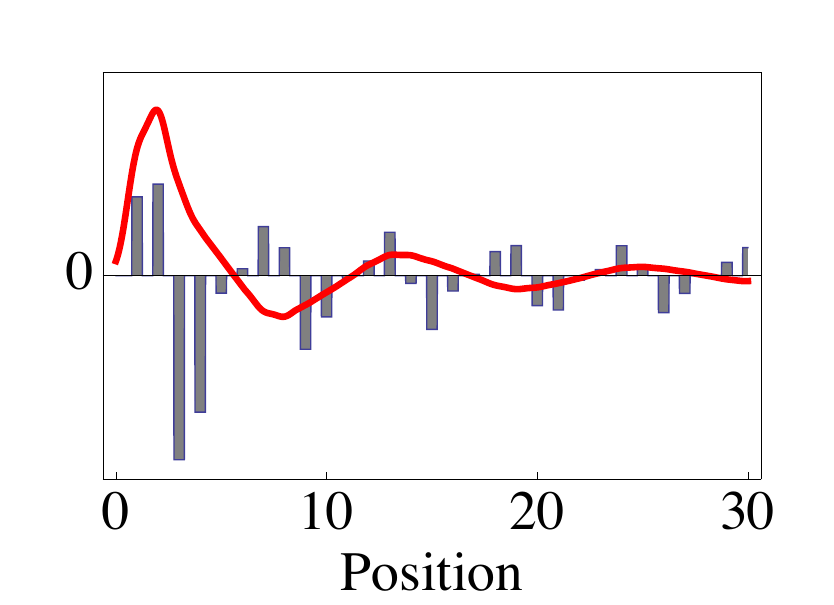}
\caption{(Color online) Example of the site energy distribution (grey bars) 
and its associated surface bound state (solid curve) with energy inside the continuum spectrum.}
\label{fig1}
\end{figure}

We consider a semi-infinite, one-dimensional linear lattice, in the presence of a site energy distribution $\{\epsilon_{n}\}$ (Fig.\ref{fig1}). In optics, this could correspond to a set of weakly-coupled optical waveguides~\cite{segev}, each of them characterized by a propagation constant $\epsilon_{n}$ and centered at $x_{n}=n a$. In the coupled-mode approach, we expand the electric field $E(x,z)$ as a superposition of the
fundamental modes centered at each waveguide, $E(x,z)=\sum_{n} C_{n}(z) \phi(x-n a)$, where $\phi(x)$ is the waveguide mode.  We pose $C_{n}(z) = C_{n} \exp(i \lambda z)$. After inserting this into the paraxial wave equation, one obtains the stationary equations for the mode amplitudes
\begin{equation}
(-\lambda + \epsilon_{n}) C_{n} + V(C_{n+1}+C_{n-1}) = 0, \\\ n>1, 
\label{eq:1}
\end{equation}
and the similar equation for the edge waveguide,
\begin{equation}
(-\lambda + \epsilon_{1}) C_{1} + V\ C_{2}  = 0, \\\\\  n =1.
\label{eq:1a}
\end{equation}
From these equations, we can formally express
\be
\epsilon_{n} = \left\{
				\begin{array}{ll}
				\lambda - V\left({C_{2}\over{C_{1}}}\right)&\mbox{if $n=1$};\\
                \lambda - V \left( {C_{n+1}\over{C_{n}}} + {C_{n-1}\over{C_{n}}} \right)& \mbox{if $n>1$}.
                \end{array}
				\right.
\label{eq:2}
\ee

For a homogeneous system, we take $\epsilon_{n}=0$, and
\be
C_{n} = \sin(k n)\hspace{1cm} \mbox{and} \hspace{1cm}\lambda = 2 V \cos(k).
\ee
For $\epsilon_{n}\neq 0$, and in the spirit of the concept of von Neumann and Wigner~\cite{wigner}, we look for 
a localized surface mode in the shape of a modulated wave of the form
\be
C_{n} = \sin(n k)\ f_{n}\label{eq:3}
\ee
where $f_{n}\rightarrow 0$ for $n\rightarrow \infty$, in order to realize a localized state. After inserting this ansatz 
into (\ref{eq:1}), we obtain
\be
\epsilon_{n} = \left\{
				\begin{array}{ll}
				\lambda - 2 V (f_{2}/f_{1}) \cos(k)&\mbox{if $n=1$}\\
                \lambda - V \{\  (f_{n+1}/f_{n}) [\cos(k)+\sin(k)\cot(k n)] \\
                + (f_{n-1}/f_{n}) [\cos(k)-\sin(k) \cot(k n)]\ \}&\mbox{if $n>1$}.
                \end{array}
                \right.		
                \label{eq:4}
\ee
and we have $\lim_{n\rightarrow\infty} \epsilon_{n}=0$, provided
that $\lim_{n\rightarrow\infty} (f_{n+1}/f_{n})=1$.

Let us take
\be
{f_{n+1}\over{f_{n}}} = (1 - \delta_{n})\label{eq:5}
\ee
where $\delta_{n}<1$.  From this, we can solve formally for $f_{n}$:
\be
f_{n} = \prod_{m=1}^{n-1}(1 - \delta_{m})\label{eq:6}
\ee
which can be rewriten as
\be
f_{n} = \exp\left\{ \sum_{m=1}^{n-1} \log(1-\delta_{m}) \right\}
\ee
In the limit $n\rightarrow\infty$, and using that $\delta_{m}<1$, we can approximate this by
\be
f_{\infty} \approx \exp\left\{ -\sum_{m=1}^{\infty}\delta_{m} \right\}
\ee
where, we want $f_{\infty}\rightarrow 0$. This implies
$\sum_{m=1}^{\infty}\delta_{m}=\infty$. A good trial function for $\delta_{n}$ is
\be
\delta_{n} = {1\over{\sqrt{n}}} \sin^{2}(n k) \sin^{2}((n+1)k).
\label{eq:delta}
\ee
The presence of the sine terms is not accidental; we need them to counteract
the presence of the two $\cot(n k)$ terms in (\ref{eq:4}) that may otherwise lead to possible divergences. In this way, we get
a smoother site energy distribution. A useful parameter to quantify the degree of localization of a state, is its participation ratio $R$, defined by, $R\equiv (\sum_{n} |C_{n}|^2)^2/\sum_{n} |C_{n}|^4$. For localized modes, $R\approx 1$ while for extended states $R\approx N$, where $N$ is the number of sites in the lattice.

\begin{figure}[h]
\includegraphics[width=0.225\textwidth]{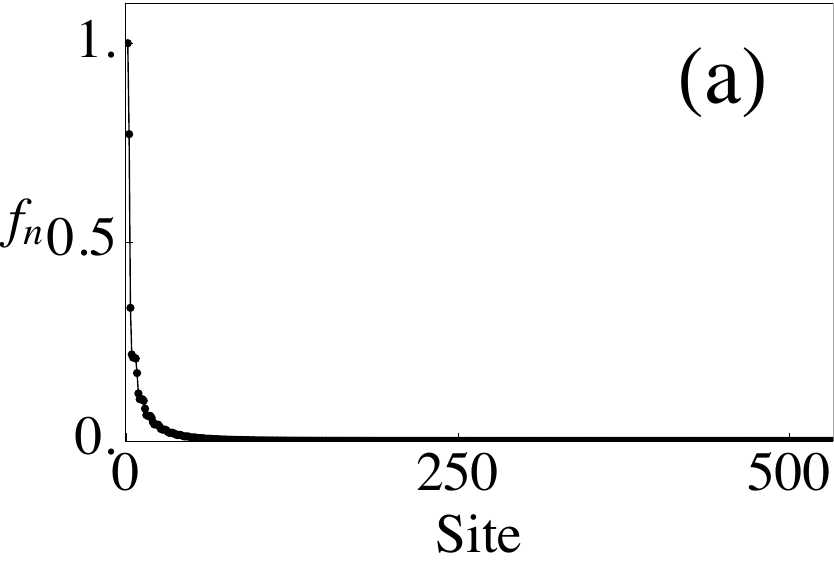}
\includegraphics[width=0.225\textwidth]{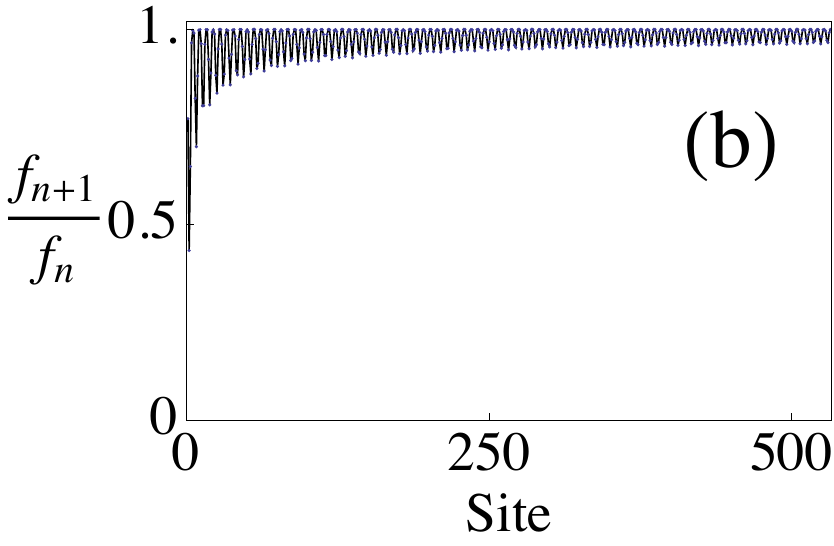}\\
\includegraphics[width=0.225\textwidth]{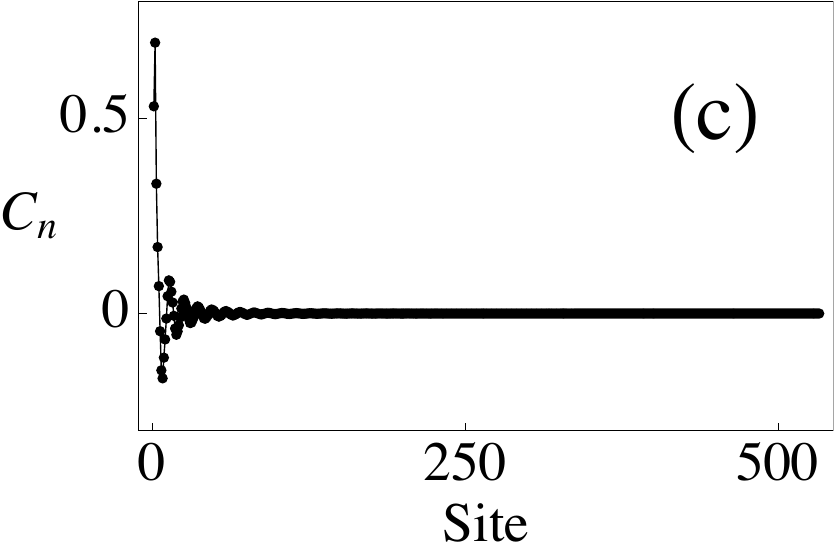}
\includegraphics[width=0.225\textwidth]{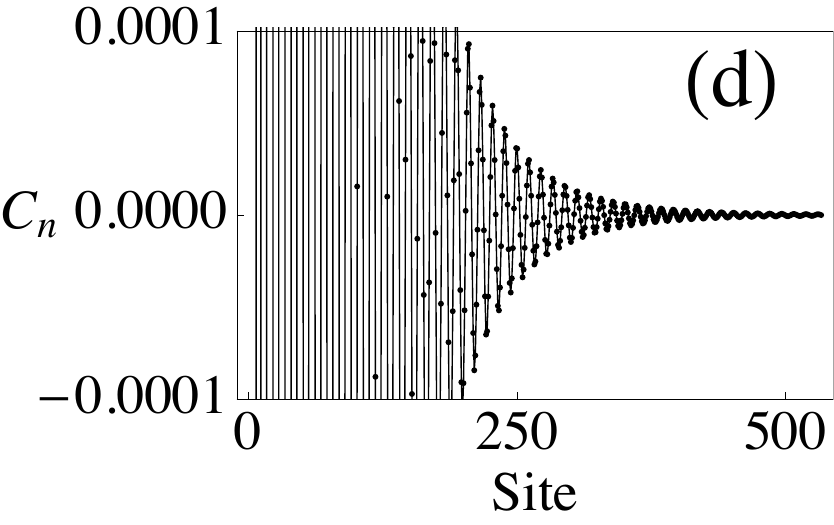}\\
\includegraphics[width=0.225\textwidth]{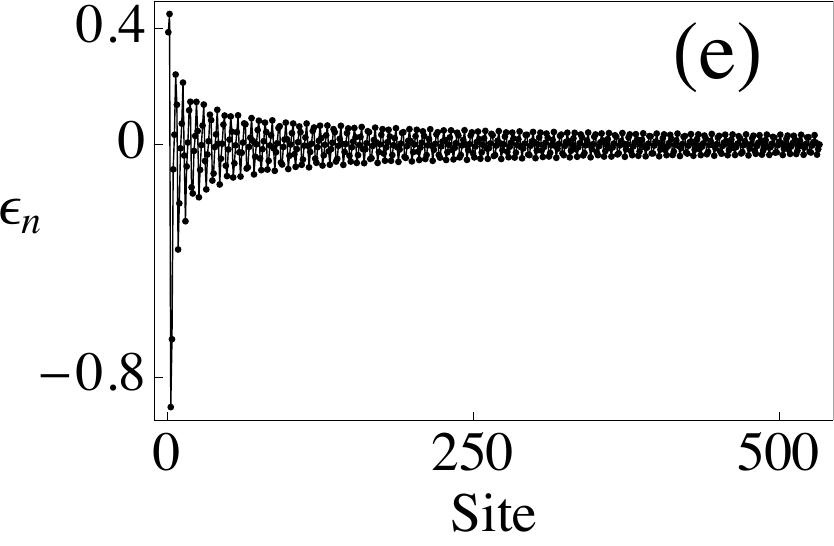}
\includegraphics[width=0.225\textwidth]{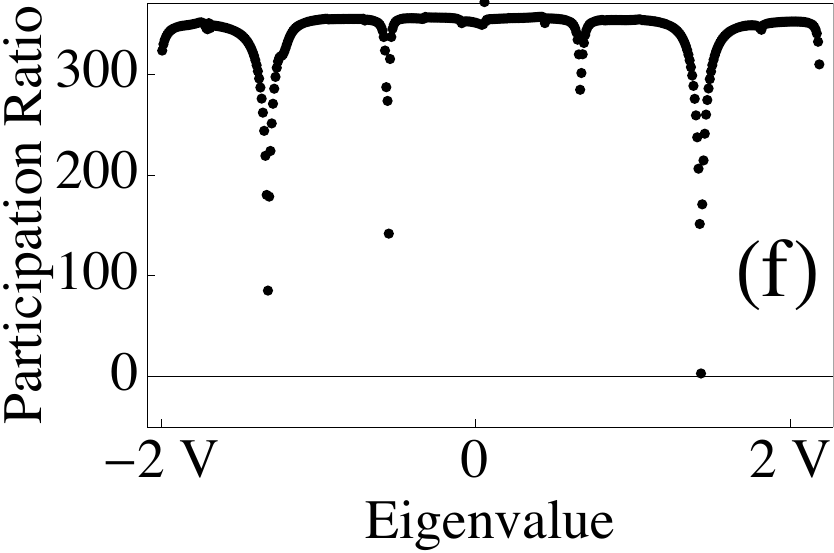}
\caption{(a) Discrete envelope function $f_{n}$ vs. $n$. (b) Ratio of the envelope
functions at the neighboring sites. (c) Embedded mode profile. (d) Close up of the panel (c).
(e) Site energy distribution and (f) participation ratio of all eigenvectors of the linear modes.
($k=0.56, N=533$).}
\label{fig2}
\end{figure}

\begin{figure}[h]
\centering
\includegraphics[width=0.45\textwidth]{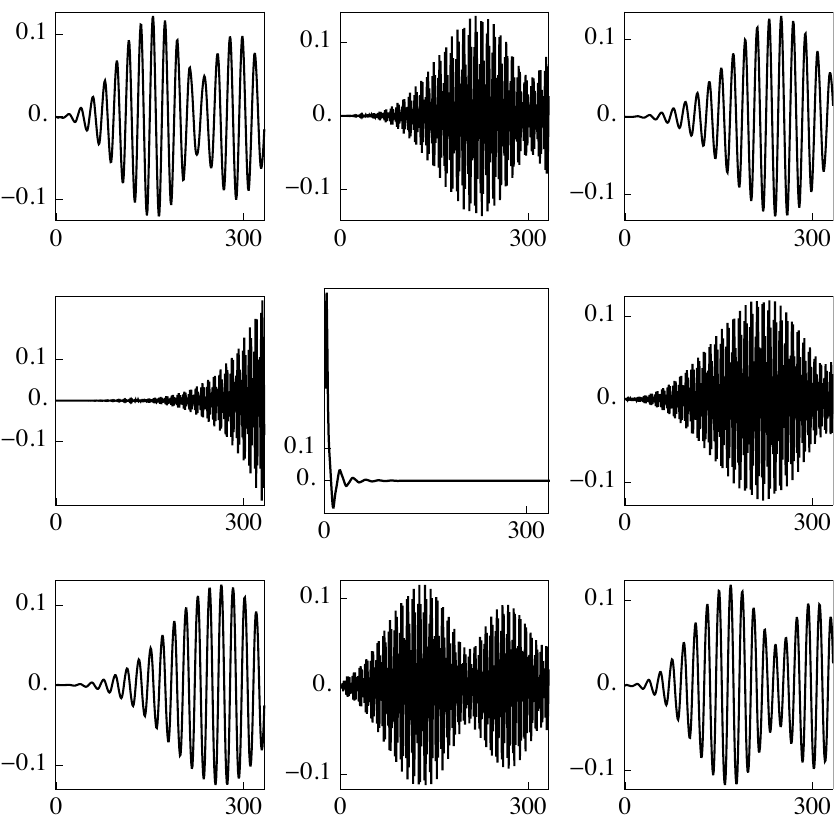}
\caption{States in the spectrum band that are the closest in energy to the embedded mode (middle panel) ($N=333, k=0.33$).}
\label{fig3}
\end{figure}

Figure \ref{fig2} shows results for a lattice of $N=533$ sites, using the trial function (\ref{eq:delta}) and $k=0.56$. The mode approaches zero slowly but surely. The asymptotic decay of the envelope at large $n$ values can be estimated, using the Euler-Maclaurin formula to be $C_{n} \sim \sin(k n) \exp(-\alpha(k) \sqrt{n})$, with $\alpha(k)=(2+\cos(2 k))/4$.
Figure \ref{fig2} also shows the site energy distribution $\epsilon_{n}$ and  participation ratio $R$ of all modes inside the band (outside the band, there are $10$ ``impurity'' localized states), and we see that our candidate for embedded mode (with eigenvalue $\lambda=1.695$) has the lowest of them all, $R\approx 3$. The next higher $R$ value is $\approx 85$. Figure \ref{fig3} shows the states inside the band that are closest in energy to the embedded state. The embedded state is the only state inside the band whose amplitude decreases to zero at large distance from the surface ($n=1$), while all the rest of the  band states are extended.

\begin{figure}[b]
\centering
\includegraphics[width=0.45\textwidth]{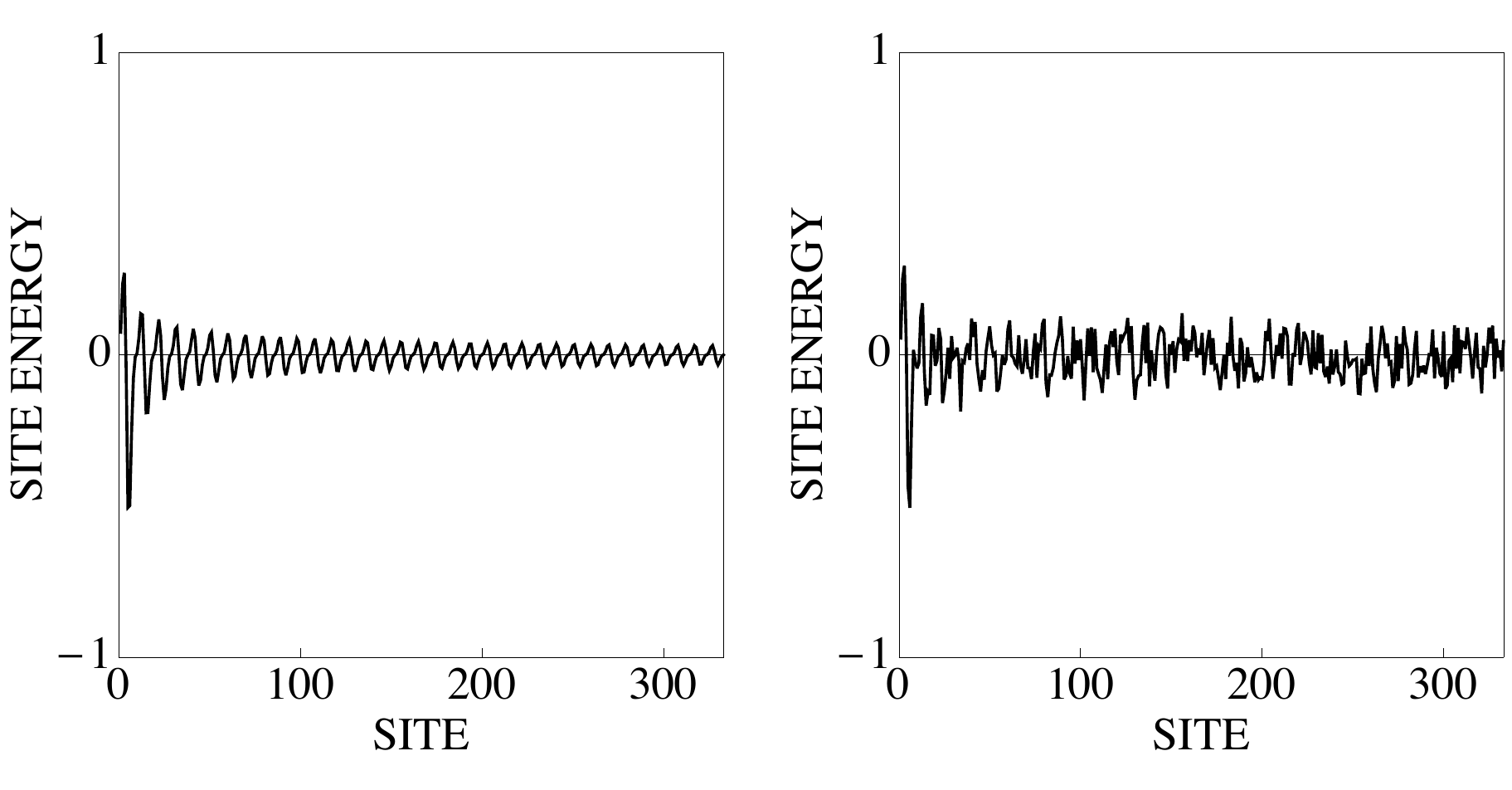}
\caption{Original (left) and randomly perturbed (right) site energy
distribution ($N=333, k=0.33$).}
\label{fig4}
\end{figure}

\begin{figure}[h]
\centering
\includegraphics[width=0.45\textwidth]{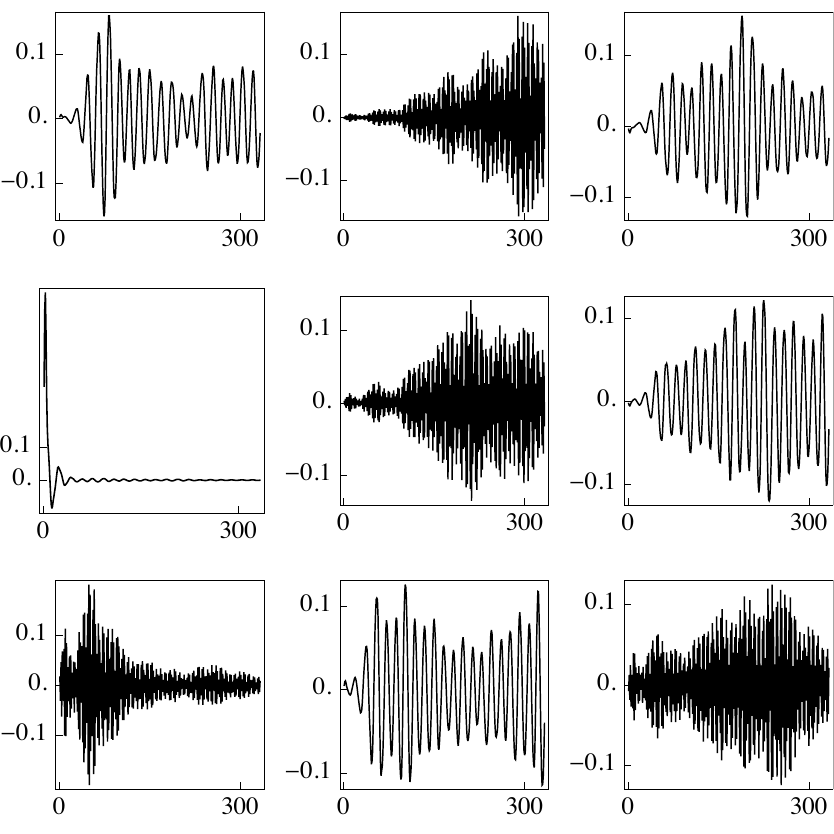}
\caption{Perturbed states in the continuum spectrum band that are the closest in energy 
to the embedded mode (left middle panel) ($N=333, k=0.33$).}
\label{fig5}
\end{figure}

Next, we proceed to check the structural stability of the embedded mode. That is, whether the mode is stable against perturbations caused, for instance by errors in the form of the site energy distribution (refraction index distribution in optics). This can happen during an attempt to reproduce experimentally the embedded mode. We took a lattice of $333$ sites and examined two cases:\\
(a) The energy site distribution
$\{\epsilon_{n}\}$ is replaced with another $\epsilon_{n}\rightarrow \epsilon_{n} + \delta_{n}$, where $\delta_{n}$ is a random number taken
from a uniform random distribution whose width is proportional to
the $\epsilon_{n}$ at a given site. For instance, we took
$\delta_{n}\in [\ -0.1\ |\epsilon_{n}^{old}|, 0.1\ |\epsilon_{n}^{old}|\ ]$.
The old and new energy site distribution look nearly the same, and as a result,
the old
given $\lambda$, just shifts a little bit. The number of states outside the  band does not change and we still have an embedded mode, surrounded by extended modes, as in Fig.~3. (b) Next, we try a more drastic perturbation, where $\delta_{n}$ is now drawn from a uniform random distribution which does not depend on site position: $\epsilon_{n}^{new}=\epsilon_{n}^{old}+\delta_{n}$ where, $\delta\in[-0.1 ,0.1]$. We see in this case that, even though the $\epsilon_{n}$ becomes significantly distorted far from the boundary, the embedded state, with eigenvalue $1.89208$ survives, with a different eigenvalue $1.85174$. Whether the new eigenvalue is smaller or large than the original one, depends on the random realization. Figure \ref{fig4} shows the old and new site energy distribution, while Fig.~\ref{fig5} shows the new band states surrounding the new embedded mode. The spatial profile of all of them maintain their extended nature, save for a small tendency towards localization, in agreement with Anderson localization theory. As long as the disorder is small and Anderson's localization length is much larger than the dimensions of the lattice, the localized embedded state is well defined. 

We can then conclude that the embedded state is structurally stable against
small perturbations.

Finally, we address the issue of possible tunability of the embedded state. Staring from a given embedded mode $\lambda$, we would like to be able to change its position inside the band, without altering the original site energy distribution. An attractive way to do this, accessible in optics, is by introducing a small amount of nonlinearity into the system. For Kerr nonlinearity, all state eigenvalues will be shifted by the amount $\gamma |C_{n}|^{2}$. For extended states, the effect will be negligible, and the only state that can be affected, is the localized one.
In the presence of a nonlinear response, the discrete equations Eqs.~(\ref{eq:1}), (\ref{eq:1a}) become:
\begin{equation}
(-\lambda+\epsilon_{n}+\gamma |C_{n}|^{2}) C_{n}+V(C_{n+1}+C_{n-1})= 0, \\\ n>1, \label{eq:12}
\end{equation}
and the equation for the edge waveguide
\begin{equation}
(-\lambda+\epsilon_{1}+\gamma |C_{1}|^{2}) C_{1}+V\ C_{2}=0, \\\ n=1,\label{eq:13}
\end{equation}
where $\gamma=1(-1)$ denotes attractive (or repulsive) nonlinearity. At this point, it is useful to make the change of variables $\phi_{n}=C_{n}/\sqrt{P}$, where $P=\sum_{n} |C_{n}|^2$ is the total power. The effective nonlinearity parameter is now $\chi=\gamma P$, and the $\phi_{n}$ are normalized to unity: $\sum_{n} |\phi_{n}|^{2}=1$. The idea is to start from the linear embedded state with given eigenvector $\lambda$ at $\chi=0$. Then, we gradually increase or decrease $\chi$ and follow the evolution of its eigenvalue and spatial profile, by solving Eqs.(\ref{eq:12}), (\ref{eq:13}) in a self-consistent manner.

\begin{figure}[t]
\centering
\includegraphics[width=0.45\textwidth]{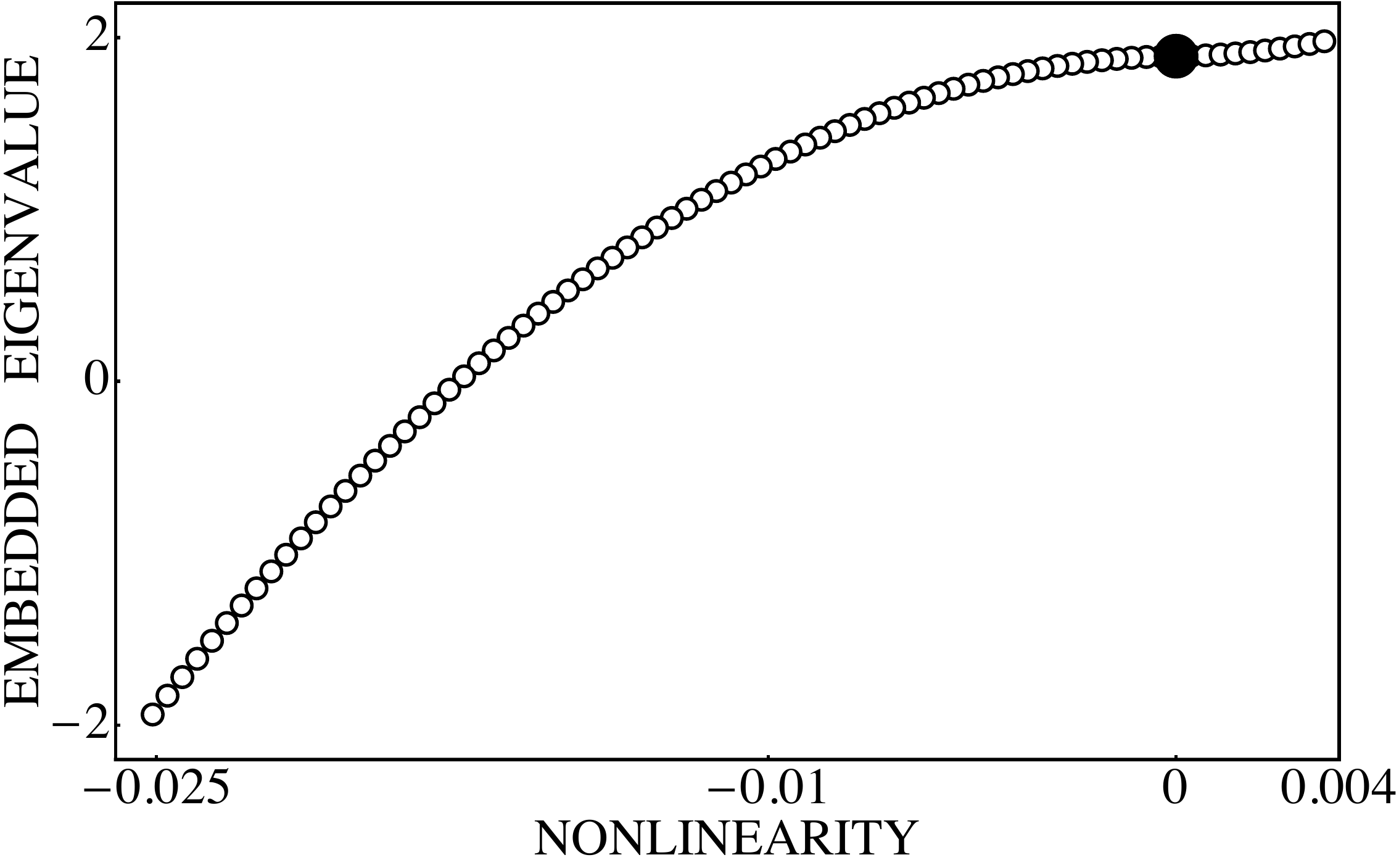}
\caption{Eigenvalue of the embedded surface mode as a function of the nonlinearity
strength, for $N=133$ and $k=0.33$. Black circle marks the position of the embedded eigenvalue in the absence of nonlinearity.}
\label{fig6}
\end{figure}

Results from this procedure are shown in Fig.~\ref{fig6}. We see that the eigenvalue of the embedded state can indeed be tuned to occur at any value inside the band, by means of a small amount of focussing or defocusing nonlinearity. The embedded state profile does not change perceptible during this process, as expected from the above discussion.

The procedure described above for one-dimensional semi-infinite discrete lattices could be generalized to higher dimensions. The simplest of such cases is a semi-infinite square lattice where the surface modes can exist near the edges and corners of a large square lattice. There, and given the separability of the Hamiltonian that gives rise to the evolution equations, the spatial profile of the embedded mode can be written as $C_{n,m} = \phi_{n} \phi_{m}$, where $\phi_{n}=\sin(n k_{x}) f^{x}_{n}$ and $\phi_{m}=\sin(m k_{y}) f^{y}_{m}$, where the envelope functions $f^{x}_{n}$ and $f^{y}_{n}$ are the same. The eigenvalue of the embedded mode is $\lambda=\lambda_{x}+\lambda_{y}$ and the energy site distribution is simply $\epsilon_{n,m}=\epsilon_{n}+\epsilon_{m}$, each is given by the appropriate modification of Eq.~(\ref{eq:4}).

In summary, we have suggested and studied a novel type of bound states localized at the edge of a semi-infinite discrete 
lattice with the eigenvalues embedded in the continuous spectrum. We have demonstrated a procedure to generate square-integrable, 
surface localized modes embedded in the continuum, as well as the site energy distributions needed to produce such modes. We have 
shown that these new embedded modes are structurally stable, and their location inside the band can be tuned by weak nonlinearity.
We believe the idea demonstrated here may be useful in other fields, including atomic systems, quantum-confined structures, 
as well as a variety of photonic structures.

The authors  acknowledge support from FONDECYT Grant 1080374 and Programa de Financiamiento 
Basal de CONICYT (FB0824/2008), and the Australian Research Council.

\end{document}